\documentclass[aps,prb,floatfix,twocolumn,superscriptaddress,showpacs]{revtex4}
\usepackage{graphicx}

\begin{document}

\title{Coulomb and Spin blockade of two few-electrons quantum dots in series in the co-tunneling regime}

\author{M. Ciorga}
\altaffiliation[Present address: ]{Experimentelle und Angewandte
Physik, University of Regensburg, D-93040 Regensburg, Germany.}
\affiliation{Institute for Microstructural Sciences,
  National Research Council of Canada,
  Ottawa, Canada K1A0R6}

\author{M. Pioro-Ladri\`{e}re}
\affiliation{Institute for Microstructural Sciences,
  National Research Council of Canada,
  Ottawa, Canada K1A0R6}
\affiliation{Centre de Recherche sur les Propri\'{e}t\'{e}s
\'{E}lectroniques de Mat\'{e}riaux Avanc\'{e}s, Universit\'{e} de
Sherbrooke, Sherbrooke, Qu\'{e}bec, J1K 2R1 Canada}

\author{P. Zawadzki}
\affiliation{Institute for Microstructural Sciences,
  National Research Council of Canada,
  Ottawa, Canada K1A0R6}

  \author{J. Lapointe}
\affiliation{Institute for Microstructural Sciences,
  National Research Council of Canada,
  Ottawa, Canada K1A0R6}

  \author{Z.Wasilewski}
\affiliation{Institute for Microstructural Sciences,
  National Research Council of Canada,
  Ottawa, Canada K1A0R6}

\author{A. S. Sachrajda}
\affiliation{Institute for Microstructural Sciences,
  National Research Council of Canada,
  Ottawa, Canada K1A0R6}

\begin{abstract}
We present Coulomb Blockade measurements of two few-electron
quantum dots in series which are configured such that the
electrochemical potential of one of the two dots is aligned with
spin-selective leads. The charge transfer through the system
requires co-tunneling through the second dot which is $not$ in
resonance with the leads. The observed amplitude modulation of the
resulting current is found to reflect spin blockade events
occurring through either of the two dots. We also confirm that
charge redistribution events occurring in the off-resonance dot
are detected indirectly via changes in the electrochemical
potential of the aligned dot.
\end{abstract}


\maketitle

\section{introduction}
Semiconductor quantum dots (QDs), often referred to as
``artificial atoms'',\cite{Kastner} have been  studied intensively
over the last decade. The spin of an electron confined within a QD
has even been suggested as a possible physical realization of a
quantum bit.\cite{Loss_qbit} Recent progress in QD related
research includes the experimental realization and study of
few-electron single\cite{ciorga00,dot_review} and
double\cite{elzerman_dd,nu2_michel} electrostatic quantum dots and
the implementation of non-local charge detectors to indirectly
detect a change in the number of confined
electrons.\cite{elzerman_dd,sprinzak}

The most common experimental transport technique used to
investigate properties of quantum dots is Coulomb blockade (CB)
spectroscopy,\cite{CB_review} from which the addition spectrum of
the system can easily be obtained. CB techniques take advantage of
the fact that to add an $N+1$ electron to a $N$-electron dot (i.e.
to observe a CB peak) one needs to match the electro-chemical
potential of the quantum dot to that of the leads. The
electrochemical potential of the dot, $\mu(N+1)$, can be tuned by
means of a ``plunger'' electrode. Since $\mu(N+1)=E(N+1)-E(N)$,
where $E(N)$ is the ground state (GS) energy of the $N$-electron
dot, any transition in the GS of either the $N$ or $N+1$ electron
dots is reflected in $\mu(N+1)$ and therefore in the plunger
electrode position of the CB peak. When the dot is connected to
spin selective leads the tunneling rates for each spin species are
different resulting in a spin blockade mechanism.
\cite{andy_spininj} Spin blockade effects provide direct
information about spin transitions through an analysis of the peak
amplitude. In particular this technique can resolve whether the
spin of two neighboring ground states differ by $+$ or $-1/2$.
However, it is important to note that this technique relies on
simultaneously measuring the property of dots with two consecutive
electron numbers, $N$ and $N+1$. For important but relatively
straightforward spin phenomena such as the `singlet-triplet'
transition for two or more electrons this will not be a
limitation. However, for more complex states such as those
involving correlations, e.g spin textures,\cite{dot_review} it is
no longer clear that states with successive occupation numbers
will overlap. In this case Coulomb blockade measurements would
automatically lead to a suppressed peak amplitude thereby
requiring a different procedure to study these novel phenomena.
Any technique which probes electron transitions at a fixed
electron number, such as the one described in this paper, may
therefore be beneficial in the future for studying these more
complex spin phenomena experimentally.

For quantum dot devices a condition of fixed $N$ is met in the
Coulomb blockaded regions when a dot is off resonance with the
leads and, as a result, the current is strongly suppressed. At low
temperatures, however, a small current still can be observed due
to higher order tunneling events. Since these involve simultaneous
tunneling of two or more electrons they are referred to as
\textit{co-tunneling}\cite{nazarov} events. For very low
temperatures and low bias voltages, including the experiments
described in this paper, co-tunneling is dominated by elastic
channels which do not involve excited electron-hole pairs within
the dot. Initially treated as a limitation on the accuracy of
single electron devices, electron co-tunneling processes have
recently been used to probe large\cite{cronenwett} and
small\cite{de_franceschi} Coulomb blockaded quantum dots as well
as Kondo systems.\cite{kondo}

In this paper we present Coulomb and spin blockade measurements of
a double quantum dot device connected to spin selective leads. In
previous experiments we used a spin polarized current to probe a
two-level molecule formed in the regime of filling factor $\nu=2$
within the dots, in which the two dots were simultaneously brought
into resonance with the leads.\cite{nu2_michel} Here, by contrast,
we focus on the situation when one of the dots is purposely kept
off-resonance. The measured current then requires a co-tunneling
process through the off-resonance dot. In order to study the
consequences of ground state transitions on the co-tunneling
current we sweep the perpendicular magnetic field close to the
well understood $\nu=2$ transition. A strong amplitude modulation
of CB peaks is observed in the co-tunneling current. The observed
modulation pattern is explained in terms of geometrical and spin
blockade effects occurring in either or both dots. The results for
the two dots in series are compared directly to measurements from
the individual dots. It is shown that this amplitude modulation
comparison can be used to identify the dominant active
co-tunneling process. In addition we observe that charge
redistributions associated with magnetic field induced ground
state transitions in the off-resonance dot, which enjoys a fixed
electron number, are reflected in changes of the electrochemical
potential of the on-resonance dot.

\section{experimental results}
A scanning electron microscope (SEM) picture of the experimental
device is shown in Fig.~1(a). The device is composed of eight
metallic gates deposited on the surface of a GaAs-AlGaAs
heterostructure with a two-dimensional electron gas (2DEG) 90 nm
below the surface. The density and mobility of 2DEG were
$n=1.7\times 10^{11} cm^{-2}$ and $\mu=2\times 10^6 cm^2
V^{-1}s^{-1}$ respectively. Individual left or right quantum dots
could be formed separately within the 2DEG by energizing different
sets of gates or alternatively a system of two few-electron dots
in series could be achieved. Plunger gates 1B and 3B were used to
tune the electrochemical potentials of the dots, and thereby the
number of electrons confined in the dots. The techniques employed
for emptying the dots and identifying the number of electrons have
been described elsewhere.\cite{ciorga00} The conductance $G$ was
measured using standard low noise lock-in techniques with a
typical bias voltage of $10\mu V$.

\begin{figure}
\includegraphics[bb=137 383 477 626,width=\columnwidth,clip]{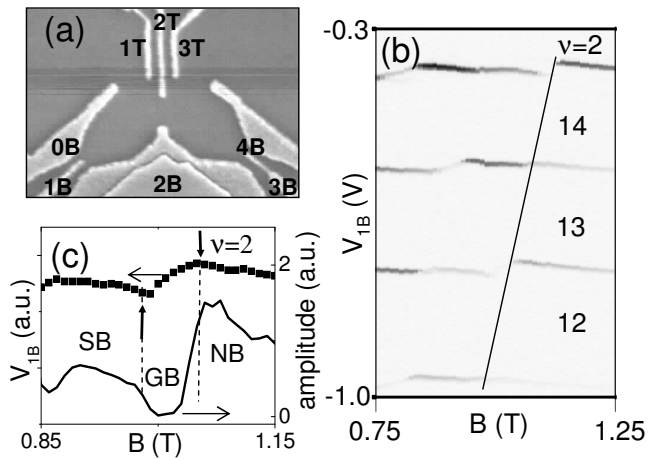}
\caption{(a) SEM picture of the experimental device; (b)inverted
greyscale of conductance through the left dot for four CB peaks in
the vicinity of $\nu=2$. Dark (light) regions represent high(low)
amplitude of CB peak; (c) typical modulation of position and
amplitude of a CB peak at $\nu=2$ transition, shown in case of the
$12\leftrightarrow 13$ transition.} \label{f1}
\end{figure}

We begin the experiment by forming the left and the right dots
individually by energizing all the gates except pairs 3B-4B and
1B-0B, respectively. To characterize each dot we map out it's
addition spectrum in a perpendicular magnetic field. We focus on
the regime near filling factor $\nu=2$ in the dot for which the
magnetic field induced ground state transitions (including spin
transitions) are very well characterized from investigations of
single quantum dot devices\cite{nu2_physE,nu2_prl,nu2_tarucha} and
the charge distribution scheme of single particle states is
particularly simple. This regime includes both spin transitions
and spatial charge redistribution events which can be used to
probe the above ideas. In Fig.~1(b) we show a typical conductance
greyscale of a single dot in the vicinity of the $\nu=2$
transition. The magnetic field dependence of four CB peaks on the
left dot is plotted. At $\nu=2$ electrons occupy a simple ladder
of states within the first Landau level (1LL) associated with an
approximately parabolic confining potential in each dot.
\cite{nu2} The wavefunction of each of these states can be
regarded as a "ring" orbital with a radius that increases with the
energy of the state. Each orbital state can be occupied by a pair
of electrons with opposite spin. Reducing the magnetic field from
it's $\nu=2$ value transfers an electron from the outermost
orbital of the 1LL to the innermost orbital of 2LL. These
transitions are reflected in the CB peak
position\cite{nu2_physE,nu2_tarucha}. For a CB peak corresponding
to the tunneling of a $N+1$ electron through an $N$-electron dot
($N\leftrightarrow N+1$ transition) both ground state transitions
of the $N+1$ and $N$ electron dot are observed as cusps. When the
$N+1$ electron tunnels through the orbital of the 1LL (2LL) the
magnetic field dependence of the position of the respective CB
peak is characterized by a downward (upward) slope. An example of
such behavior is shown in Fig.~1(c) for the $12\leftrightarrow 13$
transition. The down (up) pointing arrow identifies the GS
transition for a 13(12)-electron dot.

\begin{figure}
\label{f2}
\includegraphics[bb=86 158 509 542,width=\columnwidth,clip]{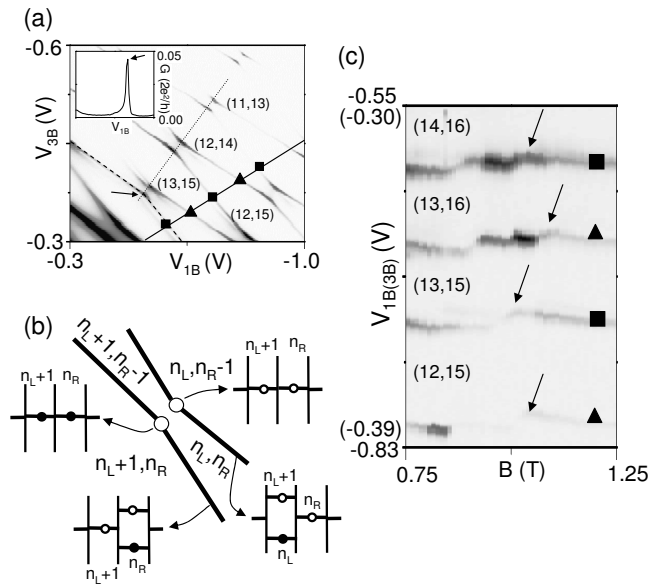}
\caption{(a) The measured charging diagram of the double dot.
Inset shows how amplitude of CB peaks rapidly decreases away from
the triple point (marked by an arrow); (b) schematic
representation of different energy configurations of a double dot
system close and away from triple points. Pair of numbers
$(n_L,n_R)$ describe occupation number for the left the and right
dot in each configuration. Open (closed) circles indicate empty
(occupied) levels in the dot; (c) inverted conductance greyscale
showing magnetic field dependence of four CB peaks, measured along
the solid line marked in (a). Solid squares (triangles) mark peaks
observed when only the left (right) dot is aligned with the leads.
Arrows mark $\nu=2$ transition as detected for each
configuration.}
\end{figure}

For an even number of electrons a spin flip is required during the
transfer of an electron between the 1LL and 2LL and a
singlet-triplet transition occurs for the $\nu=2$ dot. These spin
transitions are detected by means of spin blockade spectroscopy
and result in a strong modulation of the CB peak amplitude, as
seen in Fig.~1(b). One period of the observed amplitude modulation
is shown in Fig.~1(c) for a $12\leftrightarrow 13$ transition.
When the spin-down electron tunnels through the edge orbital of
the dot there is no blockade (NB) and a high conductance is
observed. Tunneling of spin-up electrons through the same orbital
occurs at a lower current since spin-up electrons are spin
blockaded (SB). Those two regimes are separated by a region of
very low conductance due to a different blockade mechanism. The
electron in this case tunnels through the innermost orbital,
belonging to the 2LL. Since the coupling of this orbital to the
leads is reduced due to the lateral geometry of the device we
refer to this regime as being geometrically blockaded (GB).

Two dots in series are formed by energizing all gates. Figure
~2(a) shows the greyscale of conductance vs. plunger gates
voltages. The plot, illustrating the peak positions in (1B,3B)
plane, is the measured charging diagram of the double dot (DD)
system, resembling the well-known ``honeycomb"
pattern.\cite{dd_review} The configuration within each honeycomb
is characterized by a pair of occupation numbers ($n_L, n_R$),
denoting the number of electrons in the left ($n_L$) and right
($n_R$) dot. Conductance is highest at the so-called triple
points, at which the three neighboring configurations are
degenerate. These points correspond to the situation when both
dots are in-resonance with each other and with both leads, as is
shown schematically in Fig.~2(b), so charge (current) can be
easily transferred through the system. We investigated this regime
in detail in Ref.~6 where we demonstrated the formation of
molecular states near the $\nu=2$ regime.

Let us now focus on the region away from the triple points. Along
these sides of the honeycombs only two neighboring electron
configurations are degenerate. This corresponds to the situation
where only one of the dots is in resonance with the leads. As a
result the conductance drops by over an order of magnitude, as
shown in the inset in Fig.~2(a). A voltage scan along the solid
line shown in Fig.~2(a) still results in a series of CB peaks but
with each peak corresponding to a configuration with only one dot
in resonance with the leads. Fig.~2(c) reveals the magnetic field
dependence of such a scan. Intuitively we would expect that the
current amplitude through the system would reflect the properties
of both dots (since states within both dots are involved in the
current ) but that the Coulomb blockade spectroscopy (i.e. peak
position) would only probe the on-resonance dot (since the peak
occurs when the electrochemical potential of this dot aligns with
the leads). The picture emerging from the experiment suggests this
is not the complete picture. We find three major features in the
scan: (i) two adjacent peaks are not paired in their magnetic
field behavior in contrast to peaks at the triple
points;\cite{nu2_michel} (ii) the observed amplitude modulation
reflects transitions in ground states of both in-resonance and
off-resonance dots as intuitively expected and interestingly (iii)
the peak positions reflect the transitions occurring in both dots
i.e. not only in the dot in resonance with the leads. The first of
the above observations confirms that in the regime away from the
triple points we are no longer dealing with molecular-like
behavior in the double dot system, which now behaves as two dots
in series. To analyze the second and third of the above findings
we start with a discussion of tunneling through a DD system when
only one dot is in resonance with the leads.

\section{discussion}

\begin{figure}[t]
\label{f3}
\includegraphics[bb=134 336 432 530,width=0.8\columnwidth,clip]{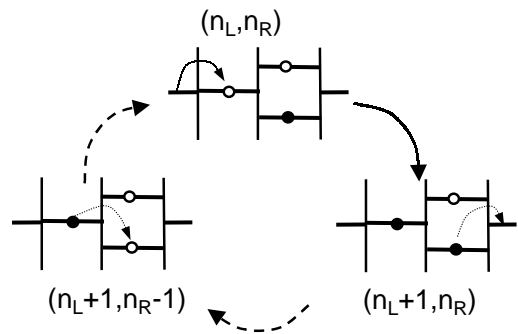}
\caption{Schematic representation of the configuration involved in
the transfer of an electron through the double dot system when
only the left dot is in resonance with the leads. Dashed arrows
indicate co-tunneling events through the virtual $(n_L+1, n_R-1)$
configuration.}
\end{figure}

Let us consider the situation when the left dot is in resonance
with the leads whereas the right dot is off-resonance. The initial
configuration is characterized by a pair of occupation numbers
$(n_L,n_R)$ and is schematically represented in Fig.~3. The $n_R$
level in the right dot lies below the Fermi energy of the leads
and so it is occupied by an electron, while the $n_L$ level in the
left dot is empty. Since the $n_L$ level is in equilibrium with
the source lead an electron can tunnel back and forth between the
source and the dot, i.e. the system configuration fluctuates
between ($n_L,n_R$) and ($n_L+1,n_R$). For the current to flow,
however, the electron needs to tunnel through the right dot. This
is possible only if we consider virtual transitions. The $n_R$
electron leaves the right dot to the lead, and a ($n_L,n_R-1$)
configuration is reached. Simultaneously, in the co-tunneling
process, the $n_L+1$ electron from the left dot enters the right
dot, the configuration ($n_L,n_R$) is achieved again, however, an
electron has been transferred through the device, i.e. current
flows. We can write down the whole cycle as
($n_L,n_R$)$\rightarrow$($n_L+1,n_R$)$\rightarrow$($n_L+1,n_R-1$)$\rightarrow$($n_L,n_R$),
where ($n_L+1,n_R-1$) is the virtual configuration accessed
through co-tunneling processes indicated in Fig.~3 by dotted
lines. We now speculate that it is possible to decompose the cycle
into the two relevant single dot transitions for the purpose of
understanding the information obtained from the amplitude
modulation. For the left dot the important transition is
$n_L\rightarrow n_L+1\rightarrow n_L$ while for the right dot it
is $n_R\rightarrow n_R-1\rightarrow n_R$. It is, of course, true
that this is a virtual process and the number of electrons in the
right dot is in reality fixed at $n_R$. However, if these
transitions are suppressed (e.g. due to spin or spatial blockade)
then the co-tunneling process will also be suppressed.

\begin{figure}
\label{f4}
\includegraphics[bb=80 101 471 725,width=\columnwidth,clip]{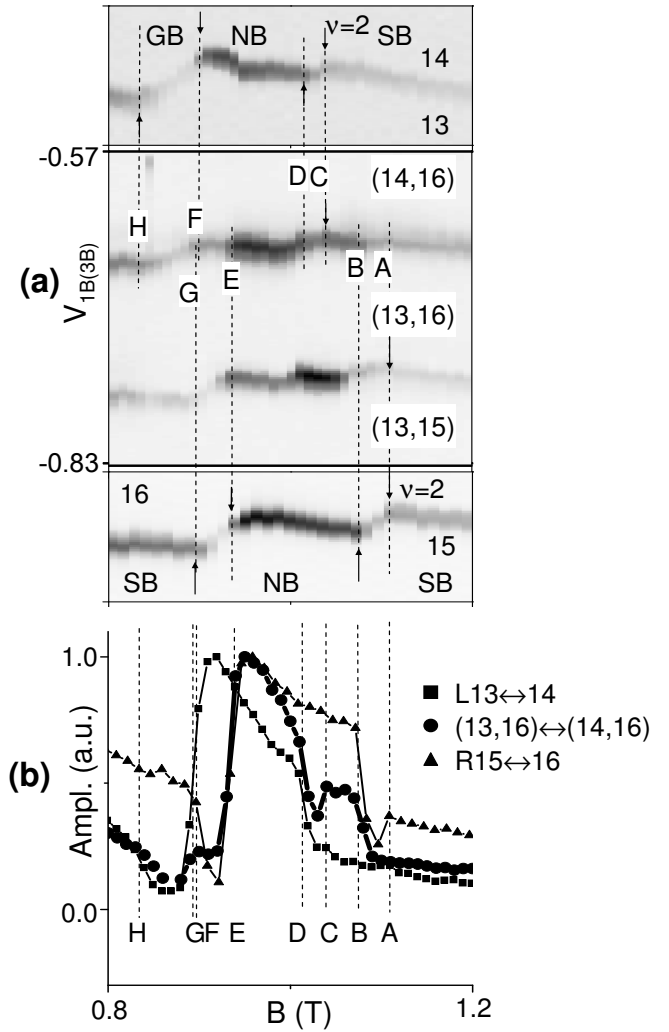}
\caption{Comparison of the double dot data for the
$(13,16)\leftrightarrow (14,16)$ with single dot data for the
$L13\leftrightarrow 14$ and the $R15\leftrightarrow 16$
transitions on the left and right dot respectively in the vicinity
of the $\nu=2$ transition. (a) respective inverted greyscales of
the double dot (middle panel), the left dot (top) and the right
dot (bottom) data. As a reference shown is also a peak related to
$(13,15)\leftrightarrow (13,16)$ transition;(b) extracted
amplitude of respective CB peaks. Arrows and dashed lines mark
transitions in ground states of respective dots, labeled by
letters for the reference. For details see text.}
\end{figure}

The above analysis is found to be consistent with our experimental
observations. As an example we choose the top peak from the
spectrum in Fig.~2(c), corresponding to the situation when
$\mu_L(14)$ is aligned with the source lead and $\mu_R(16)$ is
below the Fermi energy of the leads. According to our previous
discussion the transition $(13,16)\leftrightarrow (14,16)$ can be
written down as $(13,16)\rightarrow (14,16)\rightarrow
(14,15)\rightarrow (13,16)$ and can be decomposed into
L$13\leftrightarrow 14$ and R$16\leftrightarrow 15$ processes on
the left and the right dot respectively. Let us therefore compare
this peak with the single dot data related to transitions
$L13\leftrightarrow 14$ (left dot) and $R16\leftrightarrow 15$
(right dot). The results are shown in Fig.~4. Fig.~4(a) shows
inverted greyscale data of the relevant peaks from measurements
obtained on the double dot system (middle panel) as well as single
dot traces from the left dot (top panel) and the right dot (bottom
panel). The single dot traces have been shifted by a small amount
along horizontal axis, 40mT (60 mT) for the left (right) dot
trace, towards lower field values to match the positions of the
unambiguous $\nu=2$ transition. For reference the
$(13,15)\leftrightarrow (13,16)$ peaks is also shown. The magnetic
field values at which GS transitions occur in single dots are
marked by arrows. For a $N\leftrightarrow N+1$ transition an arrow
pointing down (up) marks a ground state transition for a $N+1
(N)$-electron dot. As described during the discussion of Fig.~1.
these transitions result in step like features in the position of
Coulomb blockade peaks with the cusps at the bottom (top) of the
steps reflecting the N (N+1) electron GS transitions. For
reference we label all of these transitions with letters $A...H$,
starting with the highest field transition, which over this range
of magnetic fields is the transition into the $\nu=2$ regime for
the 16-electron dot ($R16\leftrightarrow 15$ peak). For each peak
these transitions divide the field range into several distinctive
regions characterized by different amplitudes, which can be seen
to reflect the amplitude modulation of the relevant single dot
peaks over the same regime. The amplitude of the respective CB
peaks are shown in the Fig.~4(b). The amplitude has been
normalized to the highest value for each peak over the field
range. It is clearly seen from the data that the conductance
through the transition $(13,16)\leftrightarrow (14,16)$ is
strongest in the region $DE$ when $both$ $L13\leftrightarrow 14$
and $R16\leftrightarrow 15$ transitions are the strongest on the
individual dots i.e. when transport through the individual dots is
non-blockaded. Whenever a GS transition in one of the dots causes
a blockade mechanism to become active (either spin (SB) or
geometrical (GB)) for either of the corresponding single dot
transitions, the resulting co-tunneling current through the DD
system decreases. Lowering the magnetic field below the $E$
transition leads to decrease in current due to both GB and SB
mechanisms for the off-resonance dot (region $EF$). For magnetic
field values above the $C$ transition, regions exist where only
one of the two mechanisms is active. In the $BC$ region the
current is decreased due solely to the SB mechanism on the
in-resonance dot, but is then further reduced when the SB
mechanism is also switched on for the off-resonance dot after it
undergoes a transition to the $\nu=2$ regime (the $A$ transition).
Details of the amplitude pattern observed for all peaks from Fig.
~2(c) provide a consistent picture. Firstly, we obtain direct
information on which particular co-tunneling route is dominant. A
quite different pattern, for example, would be expected if the
co-tunneling was dominated by elastic co-tunneling through the
unoccupied state in the off-resonance dot, i.e. through
$\mu_R(17)$ in the above scenario. Secondly, a detailed analysis
of the amplitude pattern\cite{nu2_prl} confirms that spin blockade
is active in limiting the co-tunneling current.

\begin{figure}
\label{f5}
\includegraphics[bb=53 272 500 688,width=\columnwidth,clip]{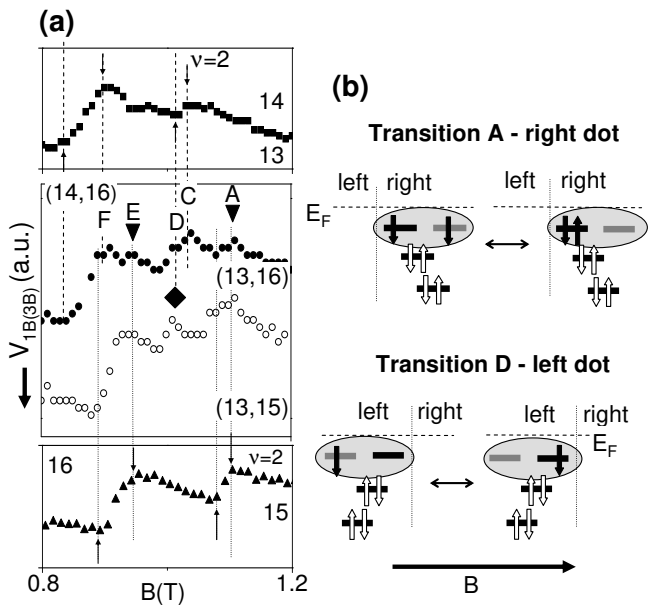}
\caption{(a) Peak positions of the data from Fig.~4(a). Curves
were offset vertically for clarity. The thick arrow indicates the
direction of higher negative plunger voltage. Ground state
transitions on single dots are marked like in Fig.~4. Black
triangles (diamond) mark features related to charge redistribution
events occurring in the right (left) off-resonance dot and picked
up by electrochemical potential of the in-resonance left (right)
dot; (b) schematic representation of ground state transitions
constituting charge redistributions. Horizontal dashed line marks
the Fermi energy of the leads ($E_F$). Vertical dotted line marks
the edge of both dots . Horizontal black (grey) bars indicate edge
(center) orbitals of the 1LL (2LL). For details see text.}

\end{figure}

Let us now discuss the peak position in the double dot traces. The
peak positions of the identical data as in Fig.~4(a) are plotted
in Fig. ~5(a). Since the left dot is always aligned with the leads
for the $(13,16)\leftrightarrow(14,16)$ peak, its position in
$V_{1B,3B}$ reflects the spectroscopy of the addition spectrum of
the left quantum dot associated with the transition
$L(13)\leftrightarrow(14)$. Similarly for the
$(13,15)\leftrightarrow(13,16)$ peak, the addition spectrum of the
right dot associated with the $R(15)\leftrightarrow(16)$
transition is measured. It can be seen that the step-like behavior
observed for single dot traces is reproduced in the double dot
data. Upward/downward cusps observed in $L(13)\leftrightarrow(14)$
($R(15)\leftrightarrow(16)$), reflecting GS transitions in 13/14
(15/16)-electron dots, align very well with the similar cusps on
$(13,16)\leftrightarrow(14,16)$ ($(13,15)\leftrightarrow(13,16)$).
It is important to note, however, that additional features are
present in the peak positions of the double dot traces. A
comparison with the single dot traces confirms that these are
related to GS transitions in the off-resonance dot. The triangles
mark the features observed in $(13,16)\leftrightarrow(14,16)$ peak
as a result of GS transitions in the off-resonance right dot with
$N=16$ electrons. One of those transition ($A$) is shown
schematically in the top panel of Fig.~5(b). As a result of this
transition a spin up electron from the outermost occupied orbital
of the 1LL (black horizontal bar) is transferred to the lowest
orbital of the 2LL (grey horizontal bar), placed close to the
center of the dot, as magnetic field is lowered. Due to the
electrostatic coupling between the dots this redistribution of
charge within the off-resonance dot causes a drop in the
electrochemical potential of the in-resonance dot shifting the
Coulomb blockade peak to more negative voltages as observed
experimentally. Similar observations are seen for other
transitions e.g. at the $(13,15)\leftrightarrow(13,16)$
transition, the charge redistribution in the off-resonance left
dot with $N=13$ (transition $D$) is picked up by the in-resonance
right dot and reflected in the spectrum (feature marked by a black
diamond in Fig.~5(a). A related observation was made recently by
Ref.~7 in the Kondo regime. This non trivial observation suggests
that integrated quantum charge detectors (dots in series or
quantum point contacts) can be used not only to detect a decrease
or increase in the number of electrons occupying the quantum dot
but also to detect rearrangements of the existing charge within a
dot. Measurements using such techniques to detect more complex
spin phenomena are currently under way.

\section{summary}
In summary, we have presented results of coulomb blockade
experiments on a double dot system with only one dot in resonance
with the leads. In this regime the current through the system is
driven by co-tunneling events through the off-resonance dots. The
co-tunneling current reflects both geometrical and spin blockade
phenomena occurring within each as well as between the two dots.
In addition we found that charge rearrangements at a fixed
electron number on one of the dots are reflected in changes of
the chemical potential of the other dot.

\begin{acknowledgments}
We would like to acknowledge useful discussions with P. Hawrylak,
M. Korkusinski, S. Studenikin and D. G. Austing. A.S. would like
to acknowledge assistance from CIAR (Canadian Institute for
Advanced Research). \end{acknowledgments}





\end{document}